# Thermal conductivity of graphene polymorphs and compounds: from $C_3N$ to graphdiyne lattices

S. Milad Hatam-Lee[1], Ali Rajabpour[1*] and Sebastian Volz[2,3*]

[1]Advanced Simulation and Computing Laboratory (ASCL), Mechanical Engineering Department, Imam Khomeini International University, Qazvin, Iran.

[2]LIMMS/CNRS-IIS(UMI2820), Institute of Industrial Science, University of Tokyo, Tokyo 153-8505 Japan.

[3]Laboratoire d'Energétique Moléculaire et Macroscopique, Combustion, UPR CNRS 288 CentraleSupélec, Université Paris-Saclay, France.

Tremendous experimental and theoretical attempts to find carbon based two-dimensional semiconductors have yielded a wide variety of graphene polymorphs, such as carbon-nitride, carbon-boride, graphyne and graphdiyne 2D materials with highly attractive physical and chemical properties. In this study, by conducting extensive non-equilibrium molecular dynamics simulations, we have calculated and compared the thermal conductivity of thirteen prominent carbon-based structures at different lengths and two main chirality directions. Acquired results show that the structures of $C_3N$, $C_3B$ and $C_2N$ exhibit the highest thermal conductivity, respectively, which suggest them as suitable candidates for thermal management systems in order to enhance the heat dissipation rates. In contrast, generally graphdiyne lattices and in particular 18-6-Gdy graphdiyne yields the lowest thermal conductivity, which can be a promising feature for thermoelectric applications. As a remarkable finding, we could establish connections between the thermal conductivity and density or Young's modulus of carbon based 2D systems, which can be employed to estimate the thermal conductivity of other polymorphs. Those results can provide a comprehensive viewpoint on the thermal transport properties of the nonporous and exceedingly porous carbon based 2D materials and may be used as useful guides for future designs in thermal management.

## 1. Introduction

Introduction of graphene and reports on its unique properties [1] motivated the synthesis of other 2D nanostructures [2]. Graphene is considered to be the most attractive 2D nanostructure that, in addition to its excellent electronic [3] and optical [4] properties, has unique mechanical [5] and thermal properties[6]. However, graphene energy bandgap is equal to zero, which limits its use in two-dimensional transistors [7], [8]. But this feature of graphene also had some positive points, including the researchers' efforts to discover and synthesize other 2D materials [9], [10]. In recent years, experimental studies on 2D semiconductor materials have brought fruitful outcomes. In this context, two-dimensional carbon-nitride nanostructures, made from covalent networks of carbon and nitrogen atoms, appeared to be among the most successful classes. For example graphitic carbon-nitride (g-C3N4) [11] with semiconducting properties was successfully synthesized and proved to be applicable in various fields including energy storage and carbon dioxide separation. [12], [13]. Other members of this family are $C_2N$ and $C_3N$ that were successfully synthesized [14], [15].

In 2015, the structure of $C_2N$ was fabricated for the first time and its energy bandgap was empirically determined [14]. In another study in 2019, the thermal properties of $C_3N$ nanotubes were investigated [16]. Recently, an experimental research has led to the synthesis of a new 2D nanostructure called N-graphdiyne with promising electronic, optical, and mechanical properties [17]. The structure of this family is very similar to that of graphyne, except that carbon atoms that bind hexagonal rings have been replaced by nitrogen atoms [18]–[20]. Shortly after, mechanical and thermal properties of $C_{18}N_6$, $C_{12}N_6$ and $C_{36}N_6$ have been calculated [18]. In an experimental work by Matsuoka *et al.* [21] in 2018, triphenylene graphdiyne (TpG) was realized experimentally. On the basis of triphenylene graphdiyne (TpG), Mortazavi et al. [22] theoretically predicted that N-triphenylene graphdiyne (N-TpG) can be a stable and strong 2D semiconducting material. Experimental successes in the discovery of the graphdiyne family have demonstrated the importance of theoretical research for understanding the intrinsic properties of materials[23]–[26].

Despite recent numerical studies on the thermal conductivity of 2D carbon-nitride and carbon-boride as well as graphdiyne structures[27]–[30], the reported results are not comparable due to the differences in the employed computational methods such as equilibrium MD, non-equilibrium MD and Boltzmann transport equation. Providing a complete picture of the thermal properties of these structures will be beneficial in the design and fabrication of new and advanced nanostructures that intend to utilize these new carbon-based 2D structures.

The purpose of the present study is to calculate and compare the thermal properties of thirteen graphene polymorphs and compounds including carbon-nitride, carbon-boride, N- graphdiyne and B- graphdiyne as shown in Fig.1 to Fig.3, utilizing non-equilibrium molecular simulations with the same potential function. Also, by replacing carbon atoms with nitrogen atoms in the $C_{18}N_6$, $C_{12}N_6$, and $C_{36}N_6$ structures, new structures are introduced that we call 18-6-Gdy, 12-6-Gdy, 36-6-Gdy and their thermal transport properties are calculated. Moreover, the effects of size and chirality on the thermal conductivity, Young's modulus and phonon mean free path of all structures are studied.

## 2. Computational details

In this study, all molecular dynamics simulations have been performed using LAMMPS [31]. The accuracy of the obtained results basically depends on the potential functions employed. In all structures, Tersoff potential function [32] with the coefficients presented by Lindsay and Brodio, is used [33]. Generally, the thermal conductivity can be calculated through equilibrium molecular dynamics (EMD) [34]–[36] or non-equilibrium molecular dynamics (NEMD) [37], [38]. In the present study, the thermal conductivity is obtained through NEMD with periodic transverse boundary conditions. In this method, a temperature gradient is first created in the system and then the resulted heat flow is measured. Thus, the thermal conductivity is deduced based on the Fourier's law relation [39]. Note that the thermal conductivity thus obtained is "effective" and includes size effects that are corrected a posteriori. Quantum population effects are also considered as negligible at room temperature due to the predominance of low frequency phonons [40].

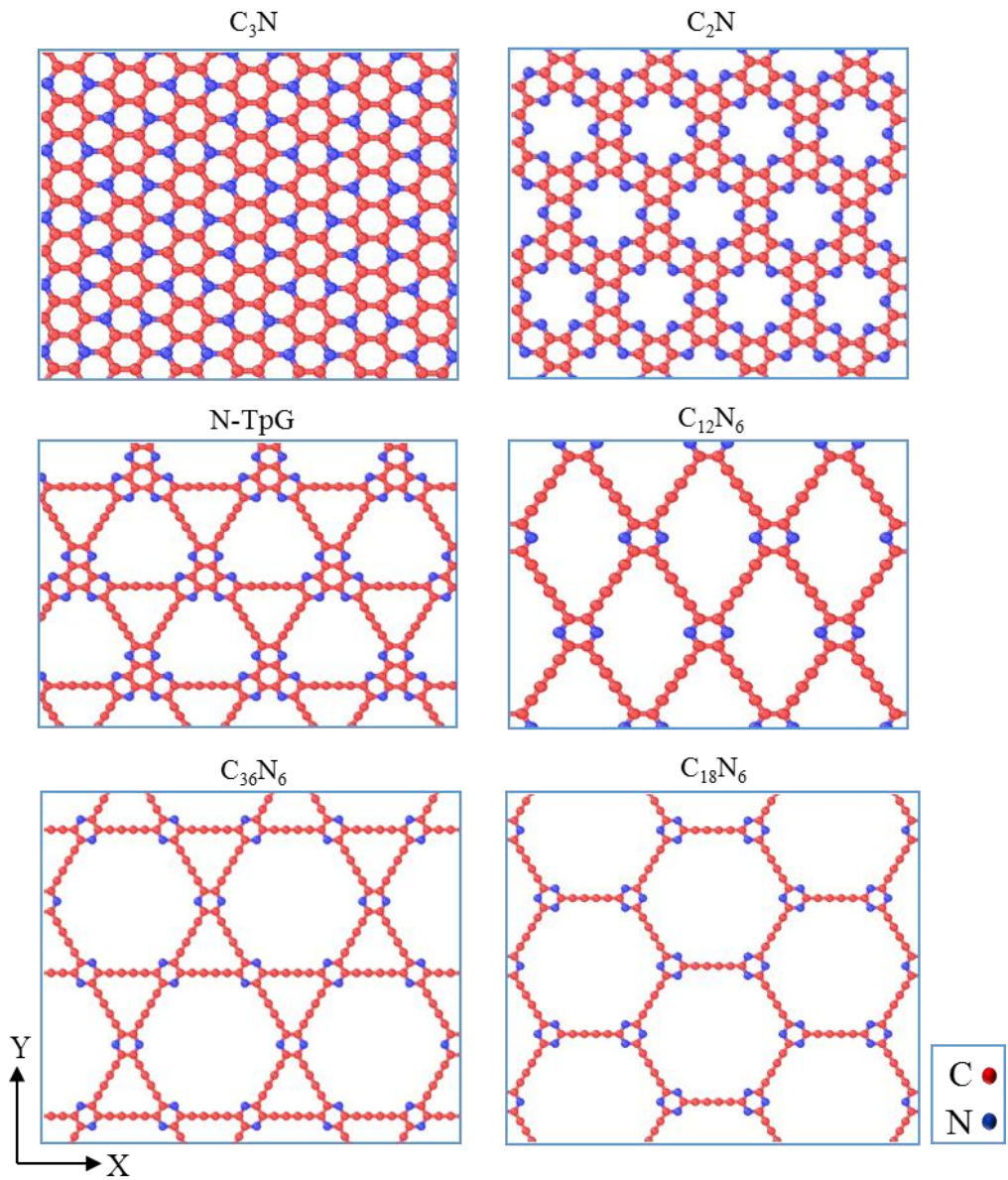

Figure 1: Atomic structure of carbon-nitride and N-graphdiyne 2D materials. Carbon and nitrogen atoms are colored red and blue, respectively.

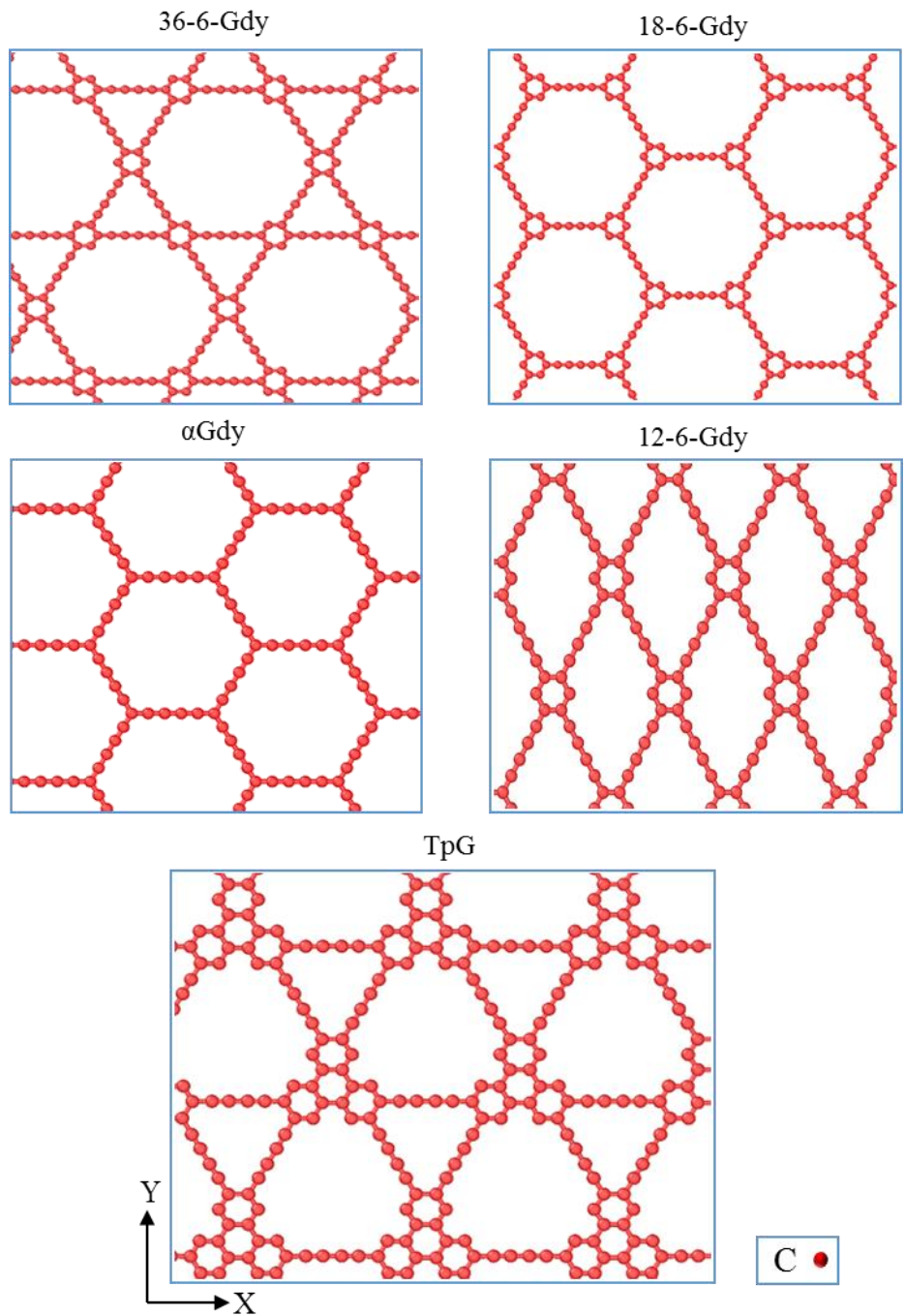

Figure 2: Atomic structure of graphdiyne 2D materials

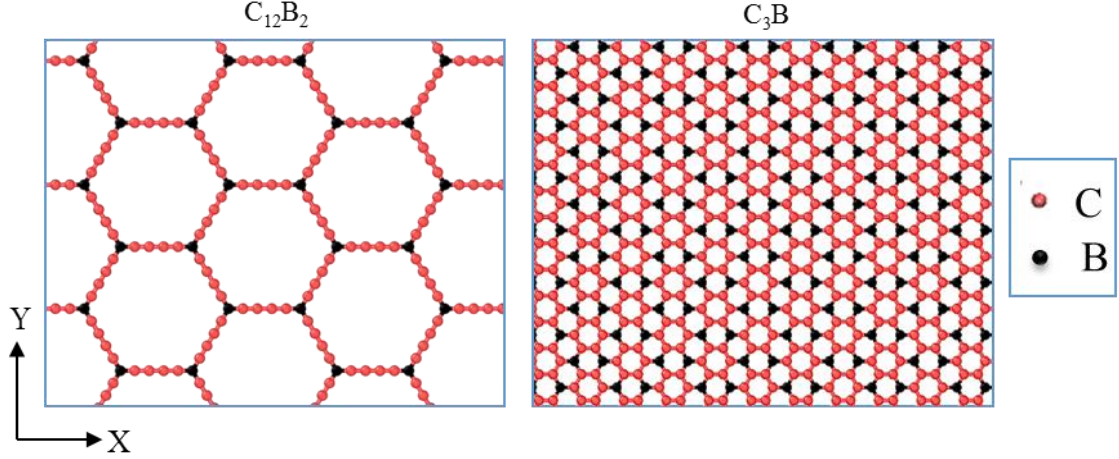

Figure 3: Atomic structure of carbon-boride ($C_3B$) and B-graphdiyne 2D materials. Carbon and boron atoms are colored red and black, respectively.

At the beginning of the simulation, the whole structure except the atoms fixed on both sides of the system as shown in Fig. 4, are thermalized up to the room temperature by Nosè-Hoover thermostat [41], [42] for 20 picoseconds. The thermostats are then switched off and the simulation box is divided into slabs (the number of slabs ranges from 27 to 42 depending on the size of the system) along the applied temperature gradient direction. The left and right ends are respectively hot and cold baths and are connected to the thermostats, while the intermediate slabs are not coupled to any thermostats. The time-step used throughout the simulation process is 0.1 femtosecond. The temperature of each slab can be obtained from the following relation:

$$T_i = \frac{2}{3 N_i k_B} \sum_j \frac{p_j}{2 m_j} \qquad (1)$$

Where $T_i$ refers to the temperature of slab i, $N_i$ is the number of atoms in the slab, $k_B$ denotes the Boltzmann constant, $m_j$ and $p_j$ are atomic mass and corresponding momentum, respectively. After thermalizing the system to the equilibrium conditions, in order to create a temperature gradient, a certain amount of kinetic energy is added to the hot bath and the same amount energy was drained through the cold bath, thus the average energy exchange between the thermostats is zero and the average temperature remains constant. After about 100 picoseconds, the heat flow and the temperature gradient in the system reach a steady-state. Over the next 4.5ns of the simulation, the thermal conductivity is obtained from the mean heat flux and the measured temperature gradient in the system.

Figure 5 (a) shows the energy diagram of heat baths. The heat flux in the X-direction is obtained from the relation $J_x = \frac{\frac{dE}{dt}}{A}$, where A is the cross-section area obtained by multiplying the width of the sheet by its thickness. The thickness of all sheets is considered to be 3.35 angstroms on average [18], [43]–[46]. Figure 5 (b) shows the temperature profiles in the system. Regardless of the temperature jumps at the left and right side of the system due to the artificial effects caused by thermostats, the temperature gradient relationship ($\frac{dT}{dx}$) throughout the sample is linear. Finally, the thermal conductivity for a sample of length L is obtained by dividing the heat flux by the temperature gradient:

$$k_L = \frac{<J_x>}{<\frac{dT}{dx}>} \tag{2}$$

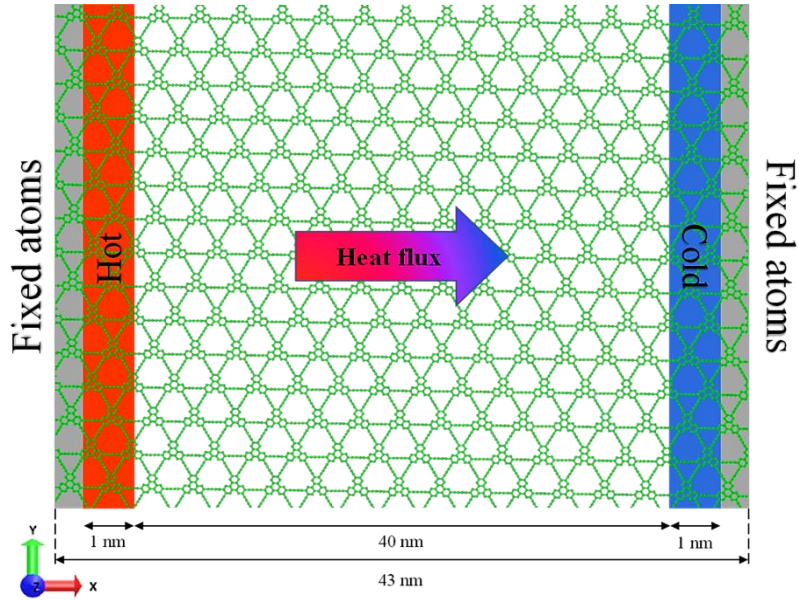

Figure 4: NEMD setup for calculating the thermal conductivity of the TpG structure.

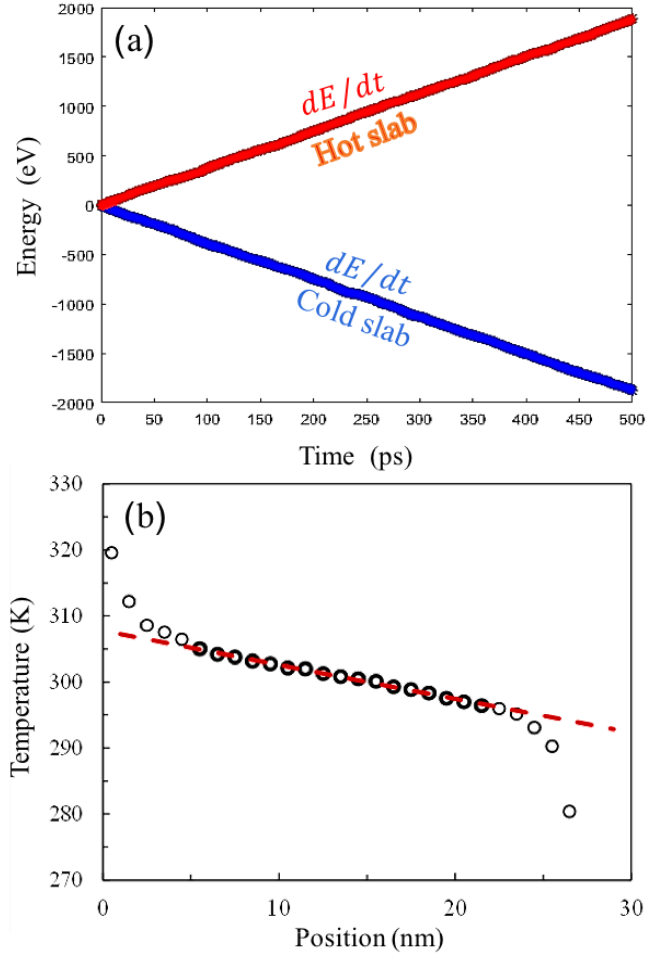

Figure 5: (a) Energy diagram for hot and cold heat baths. (b) Steady-state temperature profile in TpG structure.

## 3. Results and discussion

The investigated structures are divided into three different categories as shown in Fig. 1 to 3. Fig. 1 shows the carbon structures having nitrogen atoms, Fig. 2 illustrates the pure carbon structures and Fig.3 represent the carbon structures having boron atoms. By replacing carbon atoms with nitrogen atoms in the $C_{18}N_6$, $C_{12}N_6$ and $C_{36}N_6$ structures, three new structures have been developed that we call 18-6-Gdy, 12-6-Gdy and 36-6-Gdy. The new structures are stable at room temperature and have very similar bond lengths. The shortest and maximum C-C bond lengths in the junction of hexagonal rings are 1.38Å and 1.57Å, respectively, and within the carbon rings, the shortest and longest bond lengths are 1.24Å and 1.39Å, respectively. Using NEMD method, the calculated thermal conductivity is dependent on the size of the system due to the finite sample size which is usually smaller than the average phonon mean free path [47]. Thus the simulations were repeated for different lengths with a constant width of about 30nm at 300K.

Figures 6 to 8 shows the thermal conductivity results in two main chirality directions versus the inverse of the sample length. It has been shown that the effective thermal conductivity of a nanostructure can be represented in terms of the length using the following equation [39], [48]:

$$\frac{1}{k} = \frac{1}{k_\infty}\left(1 + \frac{\Lambda}{L}\right) \tag{3}$$

where $\Lambda$ is the effective phonon mean free path and $k_\infty$ corresponds to the thermal conductivity for the structure with infinite length. Fitting this relation into the thermal conductivity results in terms of the length, we can approximate the thermal conductivity at infinity as well as the phonon mean free path.

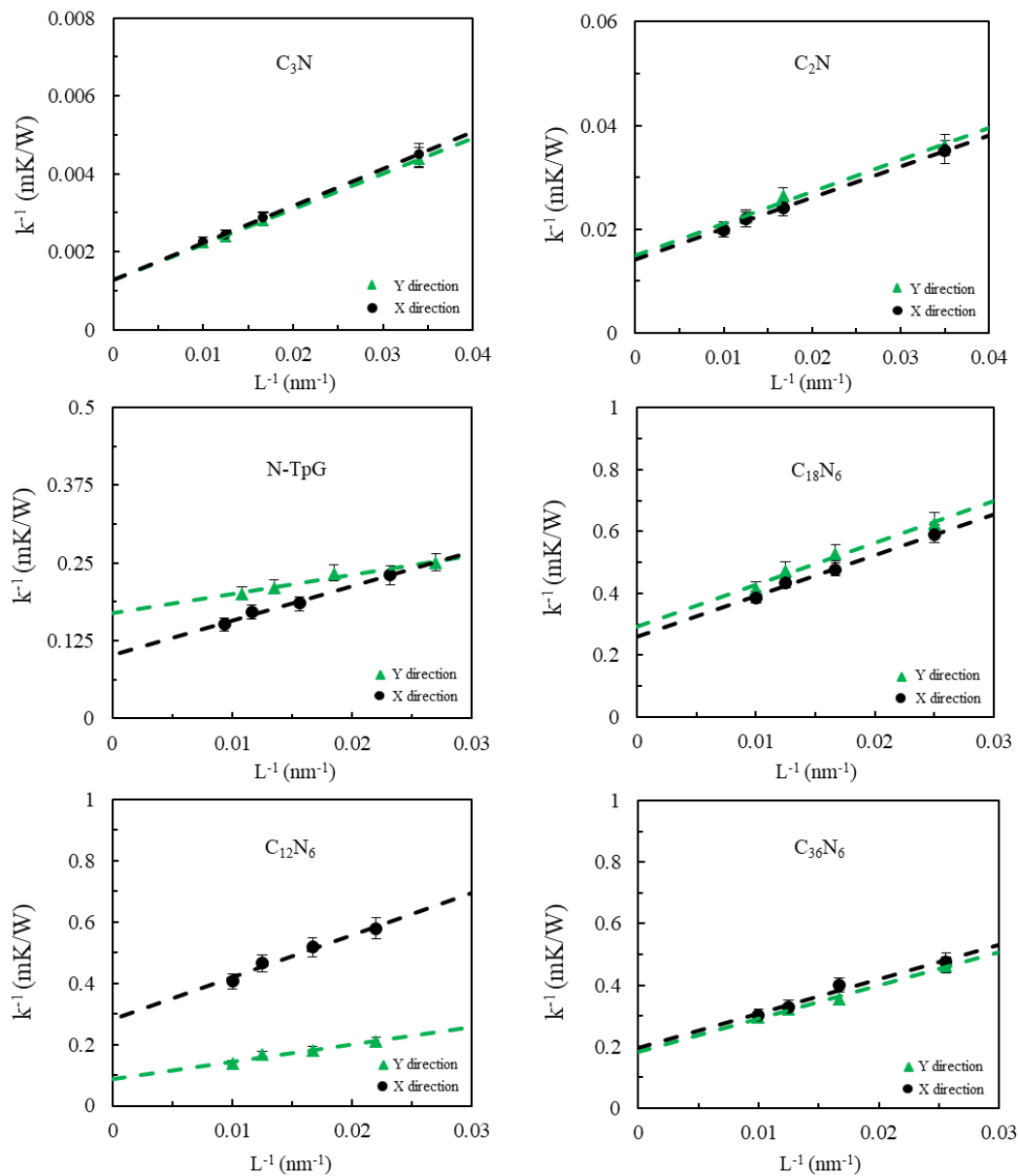

Figure 6: Thermal conductivity versus the inverse of the length for carbon-nitride and N-graphdiyne structures.

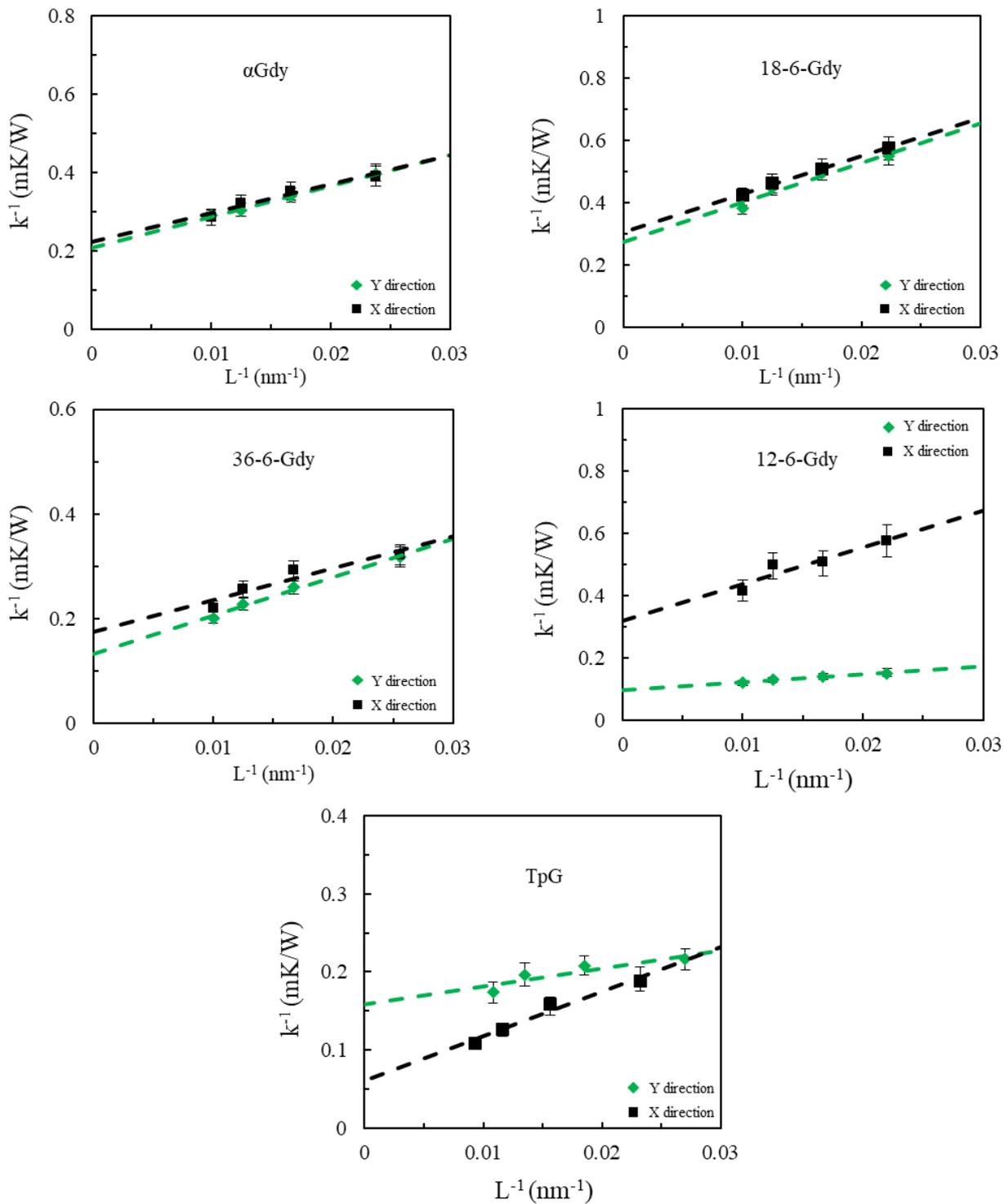

Figure 7: Thermal conductivity versus the inverse of the length for graphdiyne structures.

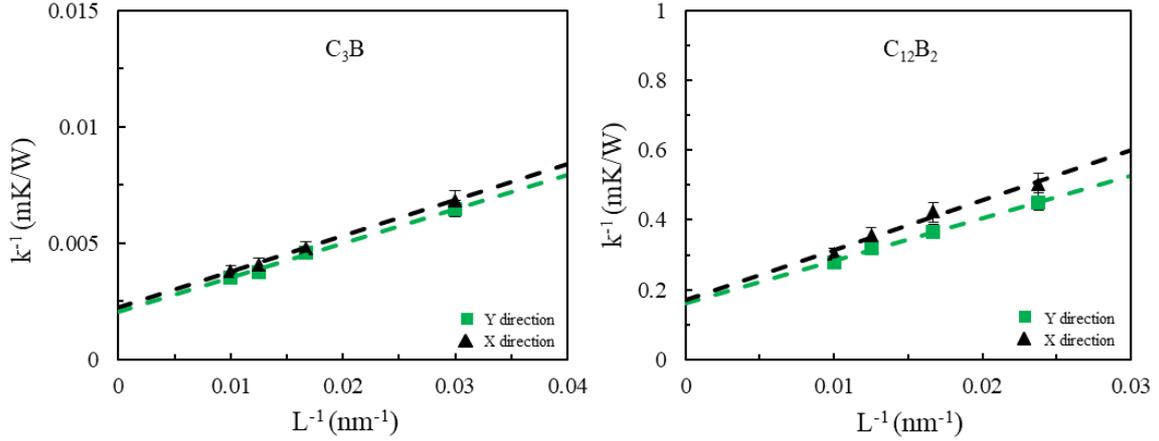

Figure 8: Thermal conductivity versus the inverse of the length for carbon-boride and B-graphdiyne structures.

The simulation results plotted in Fig. 9 to Fig. 11 show that the highest thermal conductivity about 786 W/mK belongs to the $C_3N$ structure in the X-direction, which makes it similar to graphene. This value of the thermal conductivity is in fair agreement with that of Mortazavi"s work of 2017 with 810 W/mK [49]. The thermal conductivity for $C_2N$ in the X-direction was calculated 71±4.5 W/mK is reasonably matching the value of 64.8 W/mK reported by Mortazavi et al. in 2016 [43]. Moreover, a thermal conductivity value of 451±30 W/mK was obtained for $C_3B$ structures which is close to the value reported by Song et al. (~ 488 W/mK) in 2019 using reverse NEMD method [50]. The values of $k_\infty$ for all structures in the X and Y directions are shown in Figures 9 to 11 for the carbon structures having nitrogen atoms, the pure carbon structures and the carbon structures having boron atoms, respectively.

In order to have a complete picture, Fig. 12(a) gathers the thermal conductivities of all structures for two main chirality directions in one figure. In all structures, the thermal conductivity in the two directions are approximately equal except for $C_{12}N_6$ and 12-6-Gdy, indicating the anisotropy of these two materials. Furthermore, figure 12(b) shows the thermal conductivity values sorted from the lowest to maximum for 13 structures averaged in the X and Y directions.

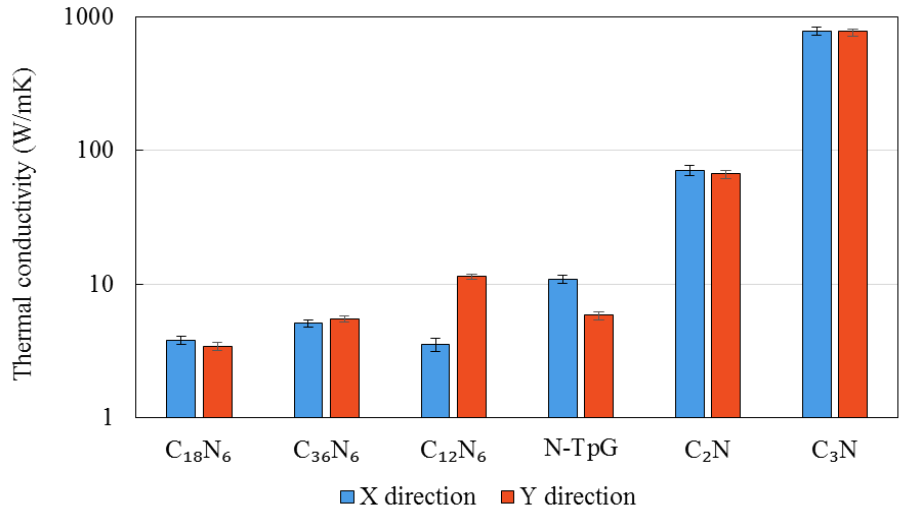

Figure 9: Thermal conductivity values at infinite length for carbon-nitride and N-graphdiyne structures.

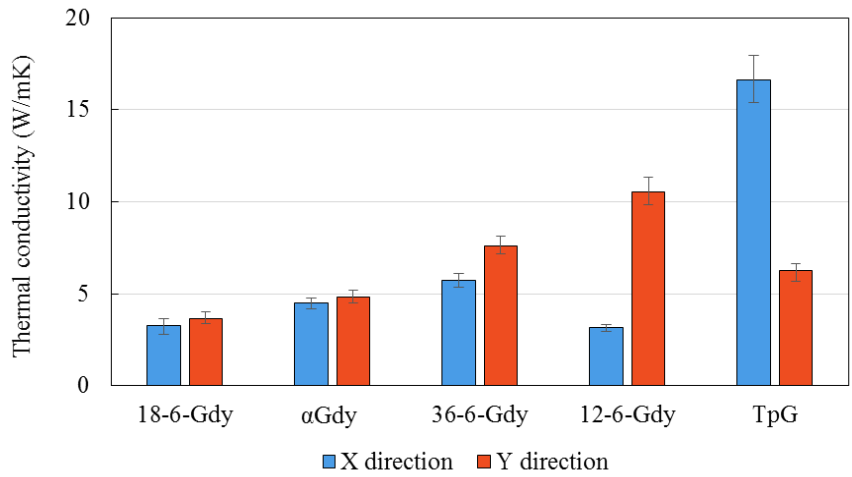

Figure 10: Thermal conductivity values at infinite length for graphdiyne structures.

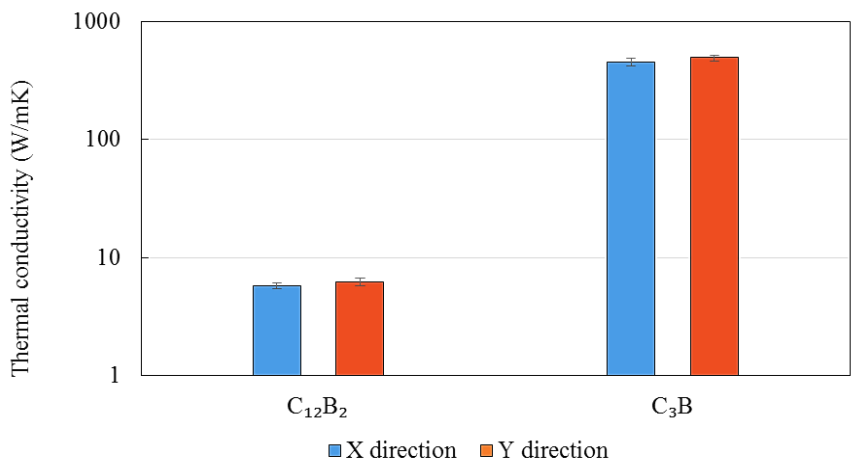

Figure 11: Thermal conductivity values at infinite length for carbon-boride and B-graphdiyne structures.

In order to better analyze the thermal transport properties of the structures, Table 1 shows the phonon mean free paths of all the materials in the X and Y directions obtained from the relation 3. As it can be seen, the mean free path varies between 14 nm to 94 nm. Although the difference in the thermal conductivity values of some structures is more than one order of magnitude, it is observed that the corresponding mean free paths are of the same order of magnitude. For instance, the two structures $C_3B$ and $C_{12}B_2$ have thermal conductivity values of 6 W/mK and 472 W/mK, but their mean free paths are 79 and 71 nm, respectively. To further investigate this mechanism of the mechanism governing heat conduction, the simplified kinetic thermal conductivity model is considered [51]:

$$k = \frac{1}{3} C_v v \, l \qquad (4)$$

where $C_v = \rho c_v$ in which $\rho$ represents the density and $C_v$ shows the heat capacity, $v$ is the average speed of sound and $l$ is the phonon mean free path. It can be seen that the thermal conductivity also depends on heat capacity and speed of sound, in addition to the dependence on mean free path. For this reason, we report these two parameters for all structures too. To calculate the heat capacity, the total energy of the structure is calculated as a function of temperature from 295 K to 305 K and heat capacity $C_v = dE/dT$ [52] is derived as the slope of this latter curve. The heat capacity for all the structures are reported in Table 1. The sound velocity is also proportional to the Young's modulus ($E_y$). To calculate Young's modulus, the stress in the structure is calculated when a longitudinal strain is applied. The slope of the stress-strain curve is equal to Young's modulus. The stress-strain curve for two structures $C_3B$ and $C_{12}B_2$ in two directions were shown in Fig.13.

In Figure 14, we compare the phonon density of states (DOS) for a few samples to the one of graphene[53] to reveal the underlying mechanism at the origin of the three orders of magnitude difference in thermal conductivity values among the carbon based 2D nanomaterials. For the case of graphene, acoustic phonons which belong to low frequency ranges are widely known as the main heat carriers. As it is clear, for the cases of highly porous and low density graphdiyne lattices, like TpG, $C_{18}N_6$ and $C_{12}B_2$, low-frequency acoustic modes contribute to the majority of phonons. This implies that the populations of these modes are substantially increased, which means that the scattering of phonons will be enhanced because the simultaneous excitation of various modes raise the rates of their collisions. In turn, the augmentation of the low-frequency mode populations results in a reduced thermal conductivity.

The PDOS of $C_3B$ and $C_3N$ are more resembling the one of graphene, which suggest lower scattering efficiency of these modes and a thermal conductivity level comparable to the one of graphene.

In Table 1, for each structure, Young's modulus and density (assuming a thickness of 0.335 nm) are presented. Young's modulus as an important parameter can indicate the mechanical strength of the structures. With the information provided in Table 1, we can provide interpretation for the difference between thermal conductivities. For instance, the $C_3B$ structure has significantly larger Young's modulus and heat capacity than that of $C_{12}B_2$ structure, indicating a large difference between the thermal conductivity of these two structures.

Moreover, in Table 1, the thermal conductivity values reported in previous research works are presented for comparison with our results. Reasonable agreements are obtained between the results of previous molecular dynamics simulations and the present work. Structures and more importantly, these can provide estimations of thermal conductivity of other carbon based 2D systems.

Table 1: Specific heat, Young's modulus, phonon mean free path and thermal conductivity in two main chirality directions.

| Material | Density $\rho$ (g/cm$^3$) | Heat capacity $c_v$ (J/cm$^3$ K) | Elastic modulus (GPa) | | Mean free path (nm) | | Thermal conductivity (W/mK) | | | |
|---|---|---|---|---|---|---|---|---|---|---|
| | | | X | Y | X | Y | X | Y | previous studies X | previous studies Y |
| 18-6-Gdy | 0.5795 | 133.2 | 128.9 | 141.6 | 40.04 | 46.54 | 3.27 | 3.65 | - | - |
| $C_{18}N_6$ | 0.6036 | 131.6 | 132 | 125 | 50.48 | 46.31 | 3.84 | 3.42 | 1.35 (EMD) [18] | - |
| αGdy | 0.7721 | 163.2 | 94.18 | 96.4 | 32.86 | 38.14 | 4.48 | 4.81 | - | - |
| $C_{36}N_6$ | 0.7787 | 170.9 | 137.8 | 154.2 | 57.18 | 58.97 | 5.11 | 5.46 | 2.5 (EMD) [18] | - |
| $C_{12}B_2$ | 0.7611 | 164.9 | 83.89 | 86.52 | 82.55 | 74.91 | 5.81 | 6.16 | 2.6 (NEMD) [42] | 2.45 (NEMD) [42] |
| 36-6-Gdy | 0.7606 | 172.9 | 81 | 98 | 35.06 | 55.76 | 5.76 | 7.59 | - | - |
| 12-6-Gdy | 1.0275 | 233.3 | 102.1 | 201.7 | 37.22 | 26.85 | 3.15 | 10.51 | - | - |
| $C_{12}N_6$ | 1.0519 | 238.0 | 119.8 | 339.1 | 49.07 | 65.08 | 3.56 | 11.48 | 1.85 (EMD) [18] | 5.75 (EMD) [18] |
| N-TpG | 1.1714 | 225.7 | 251 | 227.2 | 69.25 | 18.37 | 10.85 | 5.92 | - | - |
| TpG | 1.1337 | 244.8 | 189 | 147 | 94.92 | 14.28 | 16.61 | 6.29 | - | - |
| $C_2N$ | 1.7343 | 354.5 | 355 | 342 | 42.25 | 41.49 | 70.56 | 67.38 | 64.8 (NEMD) [55]<br>40 (EMD) [54]<br>82.22 (DFT) [45] | 40 (EMD) [54] |
| $C_3B$ | 2.1376 | 471.4 | 764 | 721 | 69.54 | 73.05 | 451.12 | 493.01 | 488 (RNME) [48]<br>412 (DFT) [56] | 489 (RNME) [48] |
| $C_3N$ | 2.2835 | 473.3 | 970 | 965 | 74.62 | 70.59 | 785.94 | 779.45 | 810 (NEMD) [47]<br>380 (DFT) [57]<br>525 (EMD) [58] | 826 (NEMD) [47] |
| Graphene | 2.267 | 480.1 | 980 | 969 | 97.42 | 99.12 | 2865.34 | 2887.47 | 3000 (NEMD) [59]<br>4000 (Exp.) [60] | - |

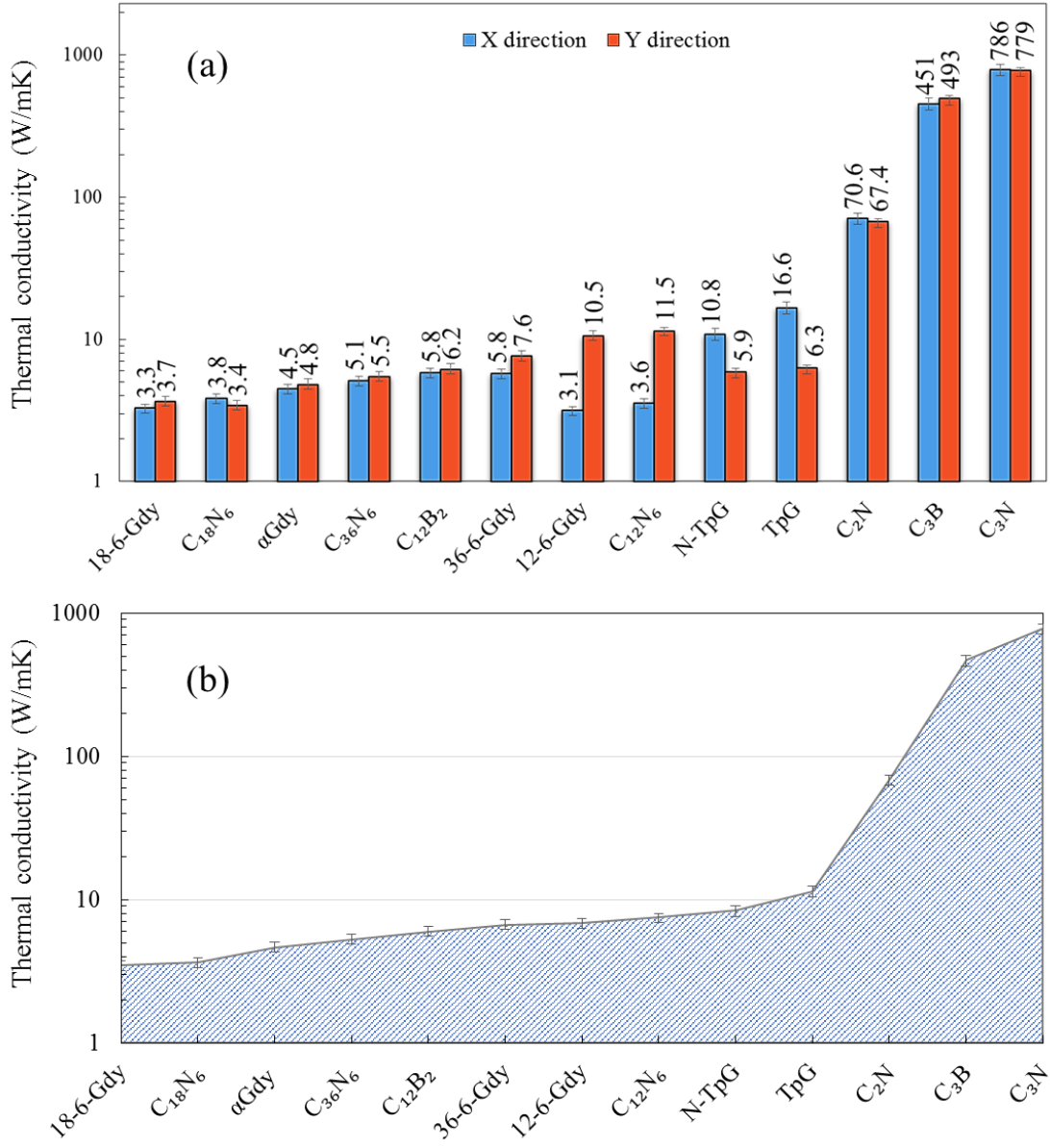

Figure 12: (a) Thermal conductivity values for thirteen carbon-based structures (graphene polymorphs and compounds) at infinite length in two main chirality directions, (b) averaged over X and Y directions.

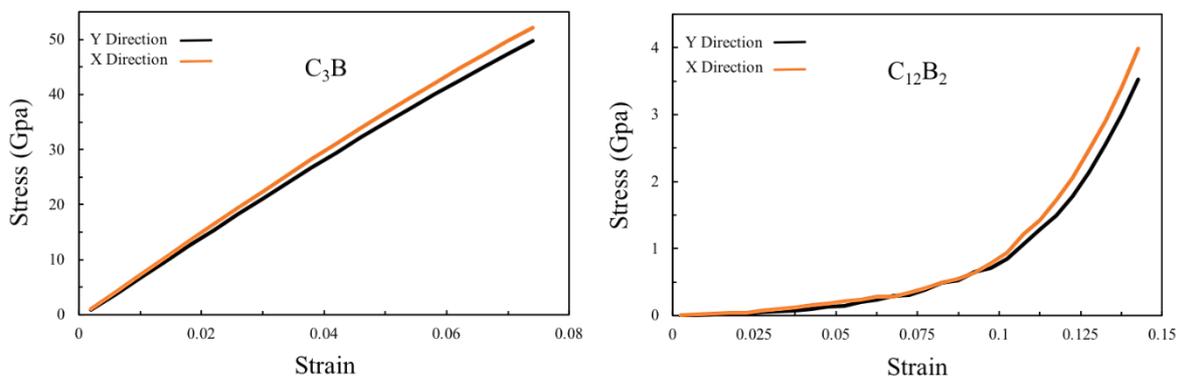

Figure 13: Stress-strain curve of $C_3B$ and $C_{12}B_2$ structures in two main chirality directions.

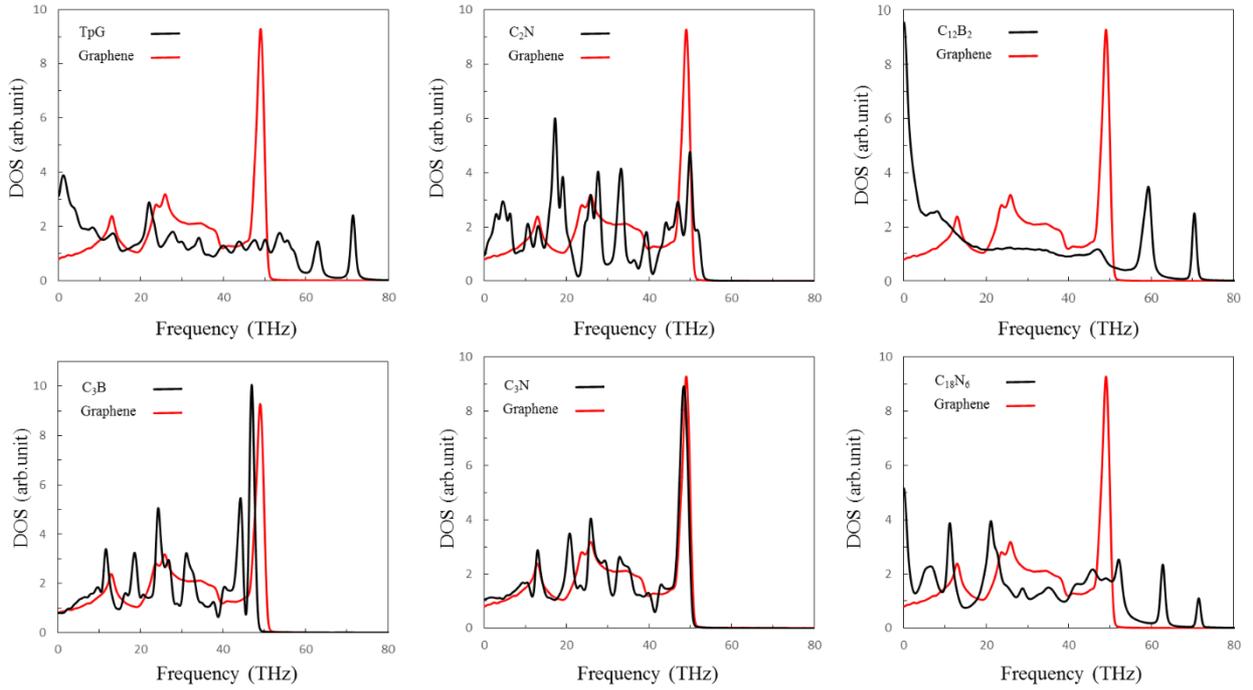

Figure 14: Phonon density of states for graphdiyne (TPG, $C_{18}N_6$ and $C_{12}B_2$), carbon-nitride ($C_2N$ $C_3N$), and carbon-boride ($C_3B$). The DOS of graphene was plotted as a benchmark.

In Table 1, the investigated structures have various densities and Young's moduli. From the first looks it is quite conspicuous that the lattices with high level of porosity not only show low rigidity but also much lower thermal conductivity. On this basis, the fundamental question is the relation between density, i.e. level of porosity, or Young's modulus and thermal conductivity. In figure 15, the thermal conductivity of all structures against density and Young's modulus are presented. . This representation reveals that exponential correlations can be established between the thermal conductivity and density or Young's modulus. Knowing these correlations provides a better understanding of thermal transport in carbon-based

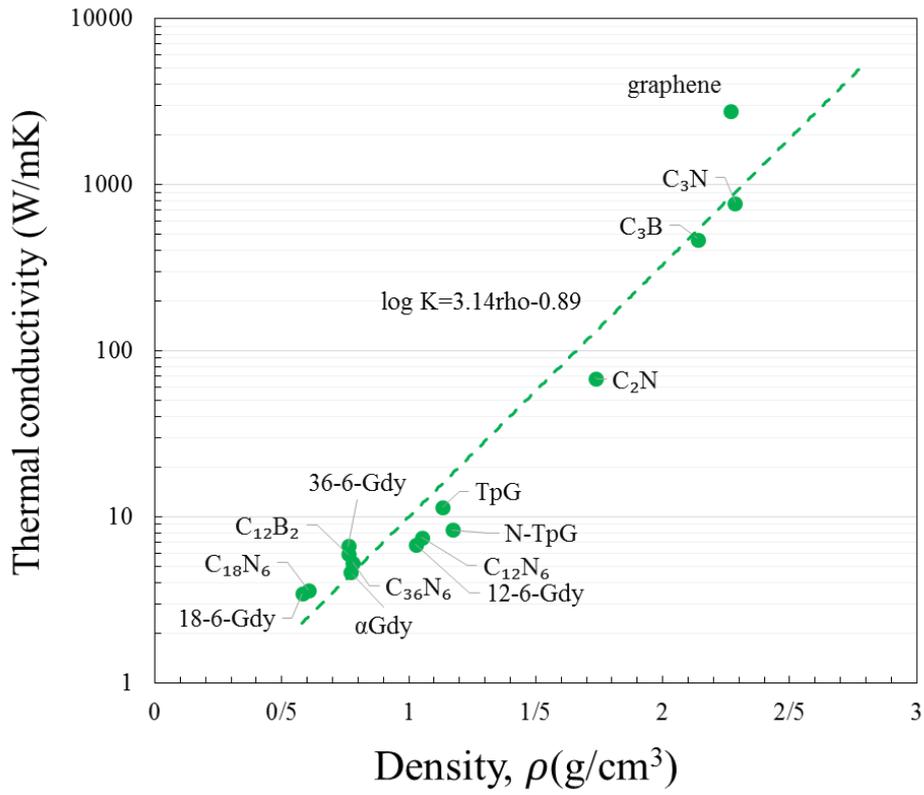

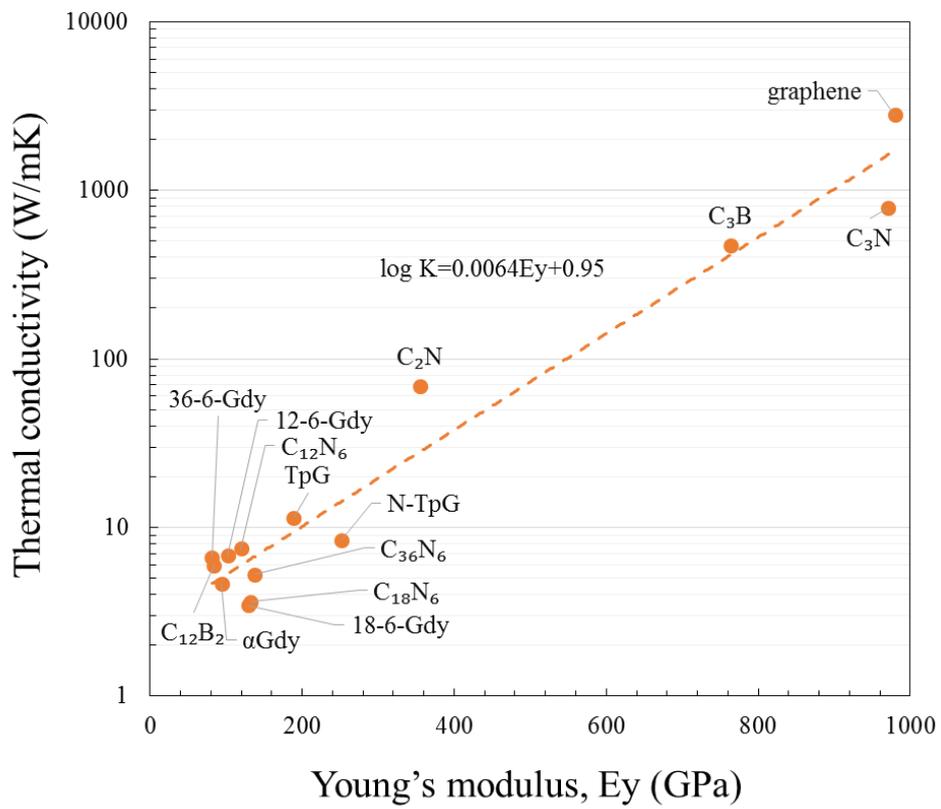

Figure 15: Thermal conductivity against density and Young's modulus.

## 4. Summary

In conclusion, by employing the non-equilibrium molecular dynamics method, the thermal conductivity of thirteen graphene polymorphs and compounds in two main chirality directions were presented. By substituting nitrogen atoms with carbon atoms in the $C_{18}N_6$, $C_{12}N_6$ and $C_{36}N_6$ structures, the new structures of 18-6-Gdy, 12-6-Gdy, 6-6-Gdy have been introduced and their thermal conduction properties have been calculated. Size dependency of the thermal conductivity was also studied for all structures and the thermal conductivity values at infinite length were estimated. Among these materials, $C_3N$, $C_3B$ and $C_2N$ were found to show the highest thermal conductivity values, while the 18-6-Gdy structure is characterized by the lowest one. The underlying mechanisms for the difference in thermal conductivity were discussed by estimating the phonon mean free path, the heat capacity and Young's modulus. Finally, we could establish connections between the thermal conductivity and the density or Young's modulus, which can be used to estimate the thermal conductivity of other nanosheets made mostly from carbon atoms. Employing the same potential function to describe atomic bonds, the MD results of this study can be a relevant benchmark for comparing the thermal conductivity of these structures. The results also provide a useful database for the thermal conductivity of these carbon-based structures for nanoelectronics applications.